\documentclass[a4paper,12pt]{article}
\usepackage{epsfig}
\date{\today}
\newcommand{\bmat}{\left(\begin{array}}
\newcommand{\emat}{\end{array}\right)}
\newcommand{\be}{\begin{equation}}
\newcommand{\ee}{\end{equation}}
\newcommand{\ba}{\begin{eqnarray}}
\newcommand{\ea}{\end{eqnarray}}

\def\lsim{\raise0.3ex\hbox{$\;<$\kern-0.75em\raise-1.1ex\hbox{$\sim\;$}}}
\def\gsim{\raise0.3ex\hbox{$\;>$\kern-0.75em\raise-1.1ex\hbox{$\sim\;$}}}

\def\be{\beta}

\def\lsim{\raise0.3ex\hbox{$\;<$\kern-0.75em\raise-1.1ex\hbox{$\sim\;$}}}
\def\gsim{\raise0.3ex\hbox{$\;>$\kern-0.75em\raise-1.1ex\hbox{$\sim\;$}}}

\begin{document}

\renewcommand{\thefootnote}{\fnsymbol{footnote}}

%\begin{titlepage}
%\pagestyle{empty}
\rightline{SUSX-TH/01-022}
\vskip 1cm
\begin{center}
{\bf \large{  Chargino Contributions to $\varepsilon$
and $\varepsilon'$\\[10mm]}}
{ Shaaban Khalil$^{1,2}$ and Oleg Lebedev$^{1}$\\[6mm]}
\small{$^1$Centre for Theoretical Physics, University of Sussex, Brighton BN1
9QJ,~~UK\\[2mm]}
\small{$^2$Ain Shams University, Faculty of Science, Cairo, 11566, Egypt \\[7mm]}
\end{center}

\hrule
\vskip 0.3cm
\begin{minipage}[h]{14.0cm}
\begin{center}
\small{\bf Abstract}\\[3mm]
\end{center}

We analyze the chargino contributions to the $K-\bar K$ mixing
and $\varepsilon'$ in the mass insertion approximation and
derive the corresponding bounds on the mass insertion parameters.
We find that the chargino contributions can significantly enlarge
the regions of the parameter space where CP violation can be fully
supersymmetric. In principle, the observed values of $\varepsilon$ and
$\varepsilon'$ may be entirely due to the chargino -- up-squark loops.

\end{minipage}
\vskip 0.7cm
\hrule
\vskip 1cm

A convenient way to parameterize SUSY contributions to the flavor changing processes
is to employ the so called mass insertion approximation \cite{Hall:1986dx}. 
The advantage of this
approach is that it allows to treat such contributions in a  model independent 
way without resorting to specific assumptions about the SUSY flavor structures
(a technical definition of this approximation will be given below).

The gluino contributions to the kaon observables in the mass insertion approximation   
have been studied in detail \cite{Hagelin:1994tc}-\cite{Ciuchini:1998ix},
but the chargino contributions have not received similar attention. The latter
have been considered either 
in the context of minimal flavor violation, that is
in SUSY models with the flavor mixing given by the CKM matrix \cite{Gabrielli:1995ff},
\cite{buras}
or as contributing to the $K-\bar K$ mixing only \cite{Branco:1994eb}.
In general, the flavor structure in the squark sector may be very complicated.
In particular, flavor patterns in the up and down sectors can be entirely different,
which may result in the dominance of the chargino contributions to the K and B 
observables.   
On the other hand, the neutralino contributions involve the same mass insertions as the  
gluino ones  (i.e. down type mass insertions) and thus cannot qualitatively change the picture.
In this Letter, we study the chargino contributions to the $K-\bar K$ mixing and $\varepsilon'$
using the mass insertion approximation and derive the  corresponding bounds on the 
mass insertion parameters.

Let us first consider the $K-\bar K$ mixing. The two observables of primary interest are
the $K_L$-$K_S$ mass difference and indirect CP violation in $K\rightarrow \pi\pi$
decays:
\begin{eqnarray}
&& \Delta M_K= M_{K_L}-M_{K_S} \;, \nonumber\\
&& \varepsilon = {A(K_L\rightarrow \pi\pi)\over A(K_S\rightarrow \pi\pi) }\;.
\end{eqnarray}
The experimental values for these parameters are $\Delta M_K \simeq 3.489 \times 10^{-15}$
GeV and $\varepsilon \simeq 2.28 \times 10^{-3}$. The Standard Model predictions for them
lie in the ballpark of the measured values, however a precise prediction cannot be made
due to the hadronic and CKM uncertainties.

Generally, $\Delta M_K$ and  $\varepsilon$ can be  calculated via
\begin{eqnarray}
&& \Delta M_K = 2 {\rm Re} \langle K^0 \vert H^{\Delta S=2}_{\rm eff} \vert \bar K^0 \rangle \;,
\nonumber\\
&& \epsilon = {1\over \sqrt{2} \Delta M_K} {\rm Im} \langle K^0 \vert H^{\Delta S=2}_{\rm eff} 
\vert \bar K^0 \rangle \;.
\end{eqnarray}
Here $H^{\Delta S=2}_{\rm eff}$ is the effective Hamiltonian for the $\Delta S=2$ transition.
It can be expressed via the Operator Product Expansion (OPE) as
\begin{equation}
H^{\Delta S=2}_{\rm eff}=\sum_{i} C_i(\mu) Q_i\;,
\end{equation}
where $C_i(\mu)$ are the Wilson coefficients and $Q_i$ are the relevant local operators.
The main uncertainty in this calculation arises from the matrix elements of $Q_i$, whereas
the Wilson coefficients can be reliably calculated at high energies and evolved down to 
low energies via the Renormalization Group (RG) running.

\begin{figure}[t]
\begin{center}
\epsfig{figure=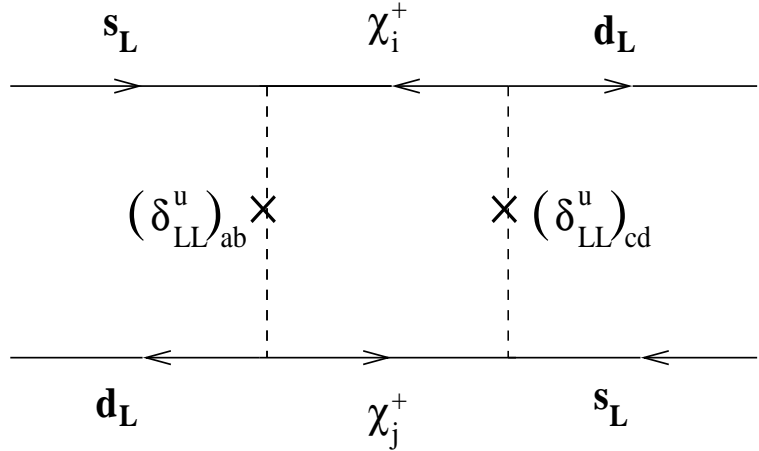,height=6cm,width=10cm,angle=0}
\medskip
\caption{Leading chargino -- up-squark contribution to $K-\bar K$ mixing. }
\label{diagram1}
\end{center}
\end{figure}

In supersymmetric extensions of the SM, the dominant chargino contribution to 
$H^{\Delta S=2}_{\rm eff}$ comes from the ``super-box'' diagram in Fig.1.
We perform our calculations in the super CKM basis, i.e. the basis in which the 
gluino-quark-squark vertices are flavor-diagonal. In this basis, the 
chargino -- left quark -- left squark vertices involve the usual CKM matrix:
\begin{equation}
\Delta {\cal L}= -g\sum_k \sum_{a,b} V_{k1} K^*_{ba}~ d_L^{a \dagger} i\sigma_2 (\tilde 
\chi^+_{kL})^* \tilde u_L^b \;, 
\end{equation}
where $K$ is the CKM matrix, $a,b$ are the flavor indices, $k=1,2$ labels the chargino
mass eigenstates, and $V,U$ are  the chargino mixing matrices defined by 
\begin{eqnarray}
&& M_{\chi^+}=\left( \matrix{ M_2 & \sqrt{2} M_W \sin\beta \cr
                           \sqrt{2} M_W \cos\beta & \mu}  \right)\;, \nonumber\\
&& U^*~M_{\chi^+}~V^{-1}={\rm diag}(m_{\chi^+_1},m_{\chi^+_2})\;.
\end{eqnarray}
Only the gaugino components of the charginos lead to significant contributions
to the $K-\bar K$ mixing since the higgsino couplings are suppressed by the 
quark masses  (except for the stop coupling) and are not important even at large $\tan\beta$.  
The stop loop contribution is suppressed by the CKM mixing at the vertices:
each vertex involving the stop is suppressed by $\lambda^2$ or $\lambda^3$ with
$\lambda$ being the Cabibbo mixing, whereas we will be working in ${\cal O}(\lambda)$
order.  The super-box involving higgsino interactions with the stops depends
on the left-right mass insertions and, as will be clear later, does not lead to
useful constraints on the SUSY flavor structures.

Due to the gaugino dominance, chargino-squark loops will generate a significant contribution
to only one operator
\begin{equation} 
Q_1= \bar s^{\alpha}_L \gamma^\mu d_L^\alpha ~  \bar s^{\beta }_L \gamma_\mu d_L^\beta\;,
\end{equation}
similarly to the Standard Model ($\alpha, \beta$ are the color indices).
The corresponding Wilson coefficient is given by the sum
of the Standard Model and the chargino contributions:
\begin{equation}
C_1(\mu)=C_1(\mu)^{SM} + C_1(\mu)^{\tilde \chi^+}\;.
\end{equation}
Generally, there are additional contributions from
gluinos and the Higgs sector, but they are not correlated with the chargino contributions
and are unimportant for the present study. We calculate $C_1(\mu)^{\tilde \chi^+}$ 
using the mass insertion approximation. That is, we express the left-left squark propagator as
\begin{equation} 
\langle \tilde u_L^a \tilde u_L^{b*} \rangle ={\rm i}~(k^2 {\bf 1}-m^2 {\bf 1}- \delta m^2)^{-1}_{ab}\simeq
{{\rm i}~\delta_{ab}\over k^2-m^2}  +{{\rm i}~(\delta m^2)_{ab}\over (k^2-m^2)^2}\;,
\end{equation}
where ${\bf 1}$ is the unit matrix and $m$ is the average up-squark mass. The SUSY contributions
are parameterized in terms of the dimensionless parameters 
$(\delta_{LL}^u)_{ab}\equiv(\delta m^2) _{ab}/m^2$. The corresponding Wilson coefficient is
calculated to be 
\begin{eqnarray}
&& C_1(M_W)^{\tilde \chi^+}={g^4\over 768 \pi^2 m^2} \left( \sum_{a,b} K^*_{a2} (\delta_{LL}^u)_{ab} K_{b1} \right)^2
\sum_{i,j} \vert V_{i1} \vert^2 \vert V_{j1} \vert^2 ~
{ x_i h(x_i)- x_j h(x_j)   \over x_i - x_j} \;,
\label{wilson}
\end{eqnarray}
where $x_i \equiv m_{\tilde \chi^+_i}^2/m^2$ and
\begin{equation}
h(x)={2+5x-x^2\over (1-x)^3} + {6x\ln x \over (1-x)^4}\;.
\end{equation}
It is interesting to note that ``flavor-conserving'' mass insertions $(\delta_{LL}^u)_{aa}$
contribute to $C_1(M_W)$, unlike for the gluino case. Such mass insertions arise from non-degeneracy
of the squark masses and are proportional to the difference of the average squark mass squared and
the diagonal matrix elements of the squark mass matrix. If the diagonal elements are equal, the
``flavor-conserving'' mass insertions drop out of the sum due to the GIM cancellations.

The flavor structure appearing in Eq.\ref{wilson} can be expanded in powers of $\lambda$:
\begin{equation}
\sum_{a,b} K^*_{a2} (\delta_{LL}^u)_{ab} K_{b1}=(\delta_{LL}^u)_{21} + 
\lambda \left[(\delta_{LL}^u)_{11}-(\delta_{LL}^u)_{22} \right] + {\cal O}(\lambda^2)\;.
\label{expand}
\end{equation}
Assuming the presence of {\it one type} of the mass insertions  at a time in Eq.\ref{expand}
at each order in $\lambda$,
one can derive constraints  on $(\delta_{LL}^u)_{21}$ and 
$\delta \equiv (\delta_{LL}^u)_{11}-(\delta_{LL}^u)_{22} $ imposed by $\Delta M_K$
and $\varepsilon$. A much weaker constraint   on $(\delta_{LL}^u)_{31}$
can also be obtained if we are to keep ${\cal O}(\lambda^2)$ terms in Eq.\ref{expand}.

To derive constraints on the mass insertions, one has to take into account the RG evolution of the
Wilson coefficients. In our numerical numerical analysis, we use  the NLO QCD result
$C_1(\mu)^{\tilde \chi^+} \simeq 0.8~ C_1(M_W)^{\tilde \chi^+} $ with $\mu=2$ GeV 
\cite{Ciuchini:1998ix}.
The matrix element of $Q_1$ is computed via 
$ \langle K^0 \vert Q_1 \vert \bar K^0 \rangle = {1\over 3} M_K f_K^2 B_1(\mu)$ with
the lattice value $B_1(\mu)=0.61$ \cite{Ciuchini:1998ix}. 
In addition,  the SM contribution should be
taken into account. Its detailed discussion  can be found in Ref.\cite{Buras:2001pn}. In our analysis,
we assume a zero CKM phase which corresponds to a conservative bound on the mass insertion.
The Wolfenstein parameters are set to $A=0.847$ and $\rho=0.4$. The other relevant constants
are $M_K=0.498$ GeV and $f_K=0.16$ GeV. 

%%% table 1  %%%
\begin{table}
\begin{center}
\begin{tabular}{|c||c|c|c|c|}
\hline
  $M_2\;\; {\huge \bf \backslash}\;\; m$   & 300 & 500 & 700 & 900 \\
\hline
\hline
150 & $0.04$ & $0.06$ & $0.08$ & $0.09$ \\
250 & $0.07$ & $0.08$ & $0.09$ & $0.11$ \\
350 & $0.09$ & $0.10$ & $0.11$ & $0.12$ \\
450 & $0.12$ & $0.12$ & $0.13$ & $0.14$ \\
\hline
\end{tabular}
\end{center}
\caption{Bounds on $\sqrt{\bigl\vert{\rm Re}\left[(\delta_{LL}^u)_{21}\right]^2\bigr\vert}$ from $\Delta M_K$ (assuming a zero CKM phase). 
To obtain the corresponding bounds on $\delta$, these entries are to be 
multiplied by 4.6. These bounds are largely insensitive to $\tan\beta$ in the
range 3--40  and to $\mu$ in the range $200-500$ GeV.}
\end{table}

The resulting bounds on $(\delta_{LL}^u)_{21}$ and
$\delta$  as functions of $M_2$ and the average squark mass $m$ are presented in Tables 1 and 2.
We find that these bounds are largely insensitive to $\tan\beta$ in the
range 3--40  and to $\mu$ in the range $200-500$ GeV. This can be understood since these
parameters do not significantly affect the gaugino components of the charginos and their
couplings. 
Note that $\delta$ is real due to the hermiticity of the squark mass matrix
and therefore does not contribute to $\varepsilon$. The presented bounds on the real part 
of $(\delta_{LL}^u)_{21}$ are a bit stronger than those derived from the 
gluino contribution to the $D-\bar D$ mixing \cite{Gabbiani:1996hi}, 
whereas the imaginary part of $(\delta_{LL}^u)_{21}$
is not constrained by any other FCNC processes.

In principle, it is possible to constrain the $(\delta_{LL}^u)_{31}$ mass insertion as well.
At order $\lambda^2$, there are two contributions  in Eq.\ref{expand}: from   
$(\delta_{LL}^u)_{31}$ and $(\delta_{LL}^u)_{12}+(\delta_{LL}^u)_{21}$. 
Assuming no cancellations between these two
terms, the  constraints on $(\delta_{LL}^u)_{31}$ are obtained by multiplying the bounds in
Tables 1 and 2 by $(A\lambda^2)^{-1}\simeq 24$. Clearly, this leaves the real part of
$\left[(\delta_{LL}^u)_{31}\right]^2$ essentially unconstrained, while the bound on 
$\sqrt{{\rm Im}\left[(\delta_{LL}^u)_{31}\right]^2}$ is of order $10^{-1}$. We note that a similar
constraint on $(\delta_{LR}^u)_{13}$ can be derived from the higgsino-stop contribution,
however such a constraint is typically satisfied automatically (especially if the squarks are heavy)
 since $(\delta_{LR}^u)_{13}
\sim \epsilon m_t/m$ with $\epsilon \ll 1$ being the 1-3 mixing the left-right sector.

%%% table 2  %%%
\begin{table}
\begin{center}
\begin{tabular}{|c||c|c|c|c|}
\hline
  $M_2\;\; {\huge \bf \backslash}\;\; m$   & 300 & 500 & 700 & 900 \\
\hline
\hline
150 & $5.3\times 10^{-3}$ & $7.2\times 10^{-3}$ & $9.1\times 10^{-3}$ & $1.1\times 10^{-2}$ \\
250 & $7.8\times 10^{-3}$ & $9.2\times 10^{-3}$ & $1.1\times 10^{-2}$ & $1.3\times 10^{-2}$ \\
350 & $1.1\times 10^{-2}$ & $1.2 \times 10^{-2}$ & $1.3\times 10^{-2}$ & $1.5\times 10^{-2}$ \\
450 & $1.5\times 10^{-2}$ & $1.5 \times 10^{-2}$ & $1.6\times 10^{-2}$ & $1.7\times 10^{-2}$ \\
\hline
\end{tabular}
\end{center}
\caption{Bounds on $\sqrt{\bigl\vert{\rm Im}\left[(\delta_{LL}^u)_{21}\right]^2\bigr\vert}$ from $\varepsilon$.
 These bounds are largely insensitive to $\tan\beta$ in the
range 3--40  and to $\mu$ in the range $200-500$ GeV.}
\end{table}

Next let us consider the chargino contribution to $\varepsilon'$ using the same approximations.
The $\varepsilon'$ parameter is a measure of direct CP violation in $K\rightarrow \pi\pi$ decays
given by
\begin{equation}
{\varepsilon'\over \varepsilon}=-{\omega \over \sqrt{2}\vert \varepsilon \vert {\rm Re}A_0}~ 
\left( {\rm Im} A_0 -{1\over \omega}~{\rm Im}A_2 \right) ,
\label{eprimeformula}
\end{equation}
where  $A_{0,2}$ are the amplitudes for the $\Delta I=1/2,3/2$ transitions and $\omega\equiv
{\rm Re} A_2 /{\rm Re} A_0 \simeq 1/22 $.  
Experimentally it has been found to be ${\rm Re}(\varepsilon'/\varepsilon)\simeq 1.9\times 10^{-3}$
which provides firm evidence for the existence of direct CP violation. This value can be accommodated
in the Standard Model although the theoretical prediction involves large uncertainties.

The effective Hamiltonian for the $\Delta S=1$ transition is given by
\begin{eqnarray}
&& H_{\rm eff}^{\Delta S=1}= \sum_i \hat C_i(\mu) \hat Q_i \;.
\end{eqnarray}
Just as in the Standard Model, two  operators, $\hat Q_6$ and $\hat Q_8$, play the dominant role.
They originate from the gluon and electroweak pengiun diagrams (Fig.2) and are defined by  
\begin{eqnarray}
&&\hat Q_6 = (\bar s^{\alpha} d^{\beta})_{V-A} \sum_{q=u,d,s} (\bar q^{\beta} q^{\alpha})_{V+A}\;,\;\nonumber\\
&&\hat Q_8 ={3\over2} (\bar s^{\alpha} d^{\beta})_{V-A} \sum_{q=u,d,s}e_q (\bar q^{\beta} q^{\alpha})_{V+A}\;,\;
\end{eqnarray}
with $(\bar f f)_{V-A}\equiv \bar f \gamma_\mu (1-\gamma_5) f$. Their matrix elements are enhanced by $(m_K/m_s)^2$
compared to those of the other operators:
\begin{eqnarray}
&& \langle (\pi \pi)_{I=0} \vert Q_6 \vert K^0 \rangle=-4 \sqrt{{3\over 2}} \left[ {m_K \over
m_s(\mu) + m_d(\mu)} \right]^2 m_K^2 (f_K-f_\pi)~B_6 \;,\nonumber\\
&& \langle (\pi \pi)_{I=2} \vert Q_8 \vert K^0 \rangle= \sqrt{3} \left[ {m_K \over
m_s(\mu) + m_d(\mu)} \right]^2 m_K^2 f_\pi ~B_8 \;, 
\end{eqnarray}
where $B_{6,8}$ are the bag parameters. In addition, the contributions of these operators are 
enhanced by the QCD corrections. Although the Wilson coefficient of $\hat Q_8$ is suppressed by 
$\alpha/\alpha_s$ compared to that of $\hat Q_6$, its contribution to $\varepsilon'$ is enhanced by 
$1/ \omega$ and is significant. In fact, it provides the dominant contribution in our analysis.

\begin{figure}[ht]
\begin{center}
\epsfig{figure=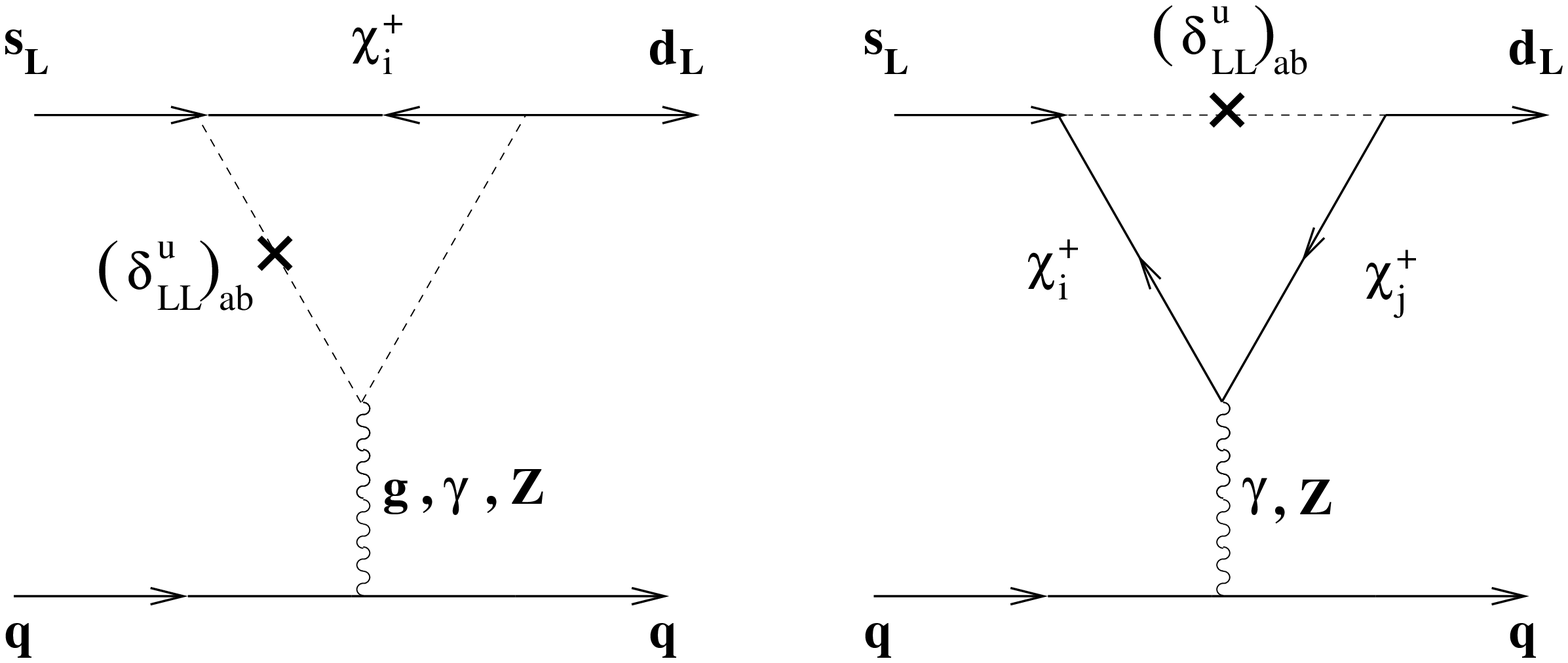,height=6cm,width=14cm,angle=0}
\medskip
\caption{Leading chargino -- up-squark contributions to $\varepsilon'$ (a ``mirror'' diagram 
is not shown). }
\label{diagram2}
\end{center}
\end{figure}

The relevant QCD corrections  in the context of the 
MSSM with minimal flavor violation have been studied in  Ref.\cite{Gabrielli:1995ff}
 and later, in more detail, in Ref.\cite{buras}.
To account for a general flavor structure in the mass insertion approximation, 
only the loop functions of Ref.\cite{buras} are to be modified.
In our numerical analysis we use the parameterization of Ref.\cite{buras} and express  the chargino contribution to 
${\varepsilon'/ \varepsilon}$ as
\begin{equation}
\left({\varepsilon'\over \varepsilon}\right)^{\tilde \chi^+}
={\rm Im} \left( \sum_{a,b} K^*_{a2} (\delta_{LL}^u)_{ab} K_{b1} \right)
\cdot F_{\varepsilon'}\;,
\end{equation}
where 
\begin{equation}
F_{\varepsilon'}= (P_X+P_Y+P_Z)~ F_Z +{1\over4}P_Z ~F_\gamma +P_E~ F_g\;. 
\end{equation}
Here we have omitted the box diagram contributions which are negligible \cite{buras}.
The parameters $P_i$ include the relevant matrix elements and NLO QCD corrections, and are given by
$P_X=0.58$, $P_Y=0.48$, $P_Z=-7.67$, and $P_E=-0.82$. The quantities  $F_i$ are functions of supersymmetric
parameters resulting from the gluon, photon, and Z penguin diagrams (Fig.2) 
and are calculated in the mass insertion
approximation. Explicitly,
\begin{eqnarray}
 F_g&=& 2  {m_W^2\over m^2} \sum_i \vert V_{i1}\vert^2 ~f_g(x_i)\;,\nonumber\\
 F_\gamma&=& 2  {m_W^2\over m^2} \sum_i \vert V_{i1}\vert^2 ~f_\gamma(x_i)\;,\nonumber\\
 F_Z&=& {1\over8} -2  \sum_i \vert V_{i1}\vert^2 f_Z^{(1)}\left({1\over x_i},{1\over x_i}\right) \nonumber\\
    &+& 2 \sum_{i,j}  V_{j1}^* V_{i1} \biggl[
U_{i1} U_{j1}^*~ f_Z^{(2)}(x_j,x_i) - V_{j1} V_{i1}^* ~f_Z^{(1)}(x_j,x_i)
\biggr]\;,
\label{formula}
\end{eqnarray} 
where $x_i \equiv m_{\tilde \chi^+_i}^2/m^2$ and the loop functions are given by
\begin{eqnarray}
&&f_g(x)= { 1-6x+18x^2-10x^3-3x^4+12x^3\ln x  \over  18(x-1)^5}\;, \nonumber\\
&&f_\gamma(x)= { 22-60x+45x^2-4x^3-3x^4+3 (3-9x^2+4x^3) \ln x\over27(x-1)^5 }\;, \nonumber\\
&&f_Z^{(1)}(x,y)={(y-1)\left[ (x-1)(x^2-x^2y+xy^2-y^2)+x^2(y-1)\ln x\right]-(x-1)^2y^2\ln y \over
                 16(x-1)^2(y-1)^2(y-x)}\;, \nonumber\\
&&f_Z^{(2)}(x,y)=\sqrt{xy} ~{(y-1)\left[ (x-1)(x-y)+x(y-1)\ln x\right]-(x-1)^2y\ln y \over
                 8(x-1)^2(y-1)^2(y-x)}\;. 
\end{eqnarray}
As noted in ref.\cite{buras}, the dominant contribution typically comes from the Z-penguin diagram, especially
if the SUSY particles are heavy. 
This can be seen as follows.
Due to the gauge invariance, the $g \bar s_Ld_L$ and $\gamma \bar s_Ld_L$ vertices are 
proportional to the second power of the momentum transfer, i.e. $(q_\mu q_\nu -g_{\mu\nu}q^2)/m^2$.
This momentum dependence is cancelled by the gluon (photon) propagator which leads to the 
suppression factor $1/m^2$ in the final result. On the other hand, the $Z \bar
s_L d_L$ vertex exists at $q^2 =0$  due to  the weak current non-conservation
and is momentum-independent to leading
order. It is given by a dimensionless function of the ratios of the SUSY
particles'  masses. The Z propagator then leads to the suppression factor
$1/M_Z^2$ which is much milder than $1/m^2$ appearing in the gluon and photon
contributions.

%%% table 3  %%%
\begin{table}
\begin{center}
\begin{tabular}{|c||c|c|c|c|}
\hline
  $M_2\;\; {\huge \bf \backslash}\;\; m$   & 300 & 500 & 700 & 900 \\
\hline
\hline
150 & $0.11$ & $0.11$ & $0.13$ & $0.16$ \\
250 & $0.17$ & $--$ & $0.87$ & $0.64$ \\
350 & $0.12$ & $0.29$ & $0.74$ & $--$ \\
450 & $0.12$ & $0.23$ & $0.42$ & $0.79$ \\
\hline
\end{tabular}
\end{center}
\caption{Bounds on $\bigl\vert {\rm Im}(\delta_{LL}^u)_{21}\bigr\vert$ from $\varepsilon'$.
For some parameter values the mass insertions are unconstrained due to the cancellations
of different contributions to $\varepsilon'$.
These bounds are largely insensitive to $\tan\beta$ in the
range 3--40; $\mu$ is set to  $200$ GeV.}
\end{table}

The resulting bounds on ${\rm Im}(\delta_{LL}^u)_{21}$ are  presented in Table 3.
Note that there is no SM contribution to $\varepsilon'$ since we assume a vanishing CKM phase.
In the limit of heavy superpartners, these bounds become insensitive to the SUSY mass scale.
This occurs due to the dominance of the Z penguin contribution.
Indeed, the contributions of the
photon and gluon penguins fall off as $1/{\rm (SUSY~ scale)^2}$ as can be seen from
Eq.\ref{formula}. On the other hand, the Z-penguin contribution stays constant in the ``decoupling'' limit.
This may seem to conflict with the intuitive expectation of the decoupling of heavy particles. However,
the proper decoupling behaviour is obtained when the flavor violating parameters $\delta m^2_{ab}$ are
kept constant (or if they grow slower than the masses of the superpartners). To put it in a slightly
different way, the decoupling should be expected when the mass splittings among the squarks grow slower
than the masses themselves. 

It is noteworthy that (for  universal GUT scale gaugino masses of around 200 GeV)
the bounds on ${\rm Im}(\delta_{LL}^u)_{21}$ are slightly stronger than those
on  ${\rm Im}(\delta_{LL}^d)_{21}$ derived from the gluino contribution to 
$\varepsilon'$ \cite{Gabbiani:1996hi}. The suppression due to the weaker coupling
is compensated by a larger loop function mainly due to the presence of the diagram
on the right in Fig.2.

These results show that to have a chargino-induced $\varepsilon'$ would require a relatively large LL
mass insertion (${\cal O}(10^{-1})$) which typically violates the constraints from
$\Delta M_K$ and $\varepsilon$. Yet, it is possible to saturate $\varepsilon$ and $\varepsilon'$
with the chargino contributions in corners of the parameter space. For instance, taking
$M_2=450$ GeV and $m=300$ GeV, $\varepsilon$ requires 
\begin{equation}
2~{\rm Im}(\delta_{LL}^u)_{21}~{\rm Re}(\delta_{LL}^u)_{21} \simeq 2.3 \times 10^{-4}\;.
\end{equation}
Then, assuming ${\rm Im}(\delta_{LL}^u)_{21}\simeq0.12$ to produce $\varepsilon'$,
${\rm Re}(\delta_{LL}^u)_{21}$ has to be $9\times 10^{-4}$. These values are in marginal
agreement with the $\Delta M_K$ bound:
\begin{equation}
\sqrt{\Bigl\vert \left[{\rm Re}(\delta_{LL}^u)_{21}\right]^2 - \left[{\rm Im}(\delta_{LL}^u)_{21}\right]^2   \Bigr\vert} \leq 0.12\;. 
\end{equation}

The main lesson, however,  is that
combining the chargino and the gluino contributions can provide fully supersymmetric
$\varepsilon$ and $\varepsilon'$ in considerable regions of the parameter space\footnote{
In principle it is possible to saturate both $\epsilon'$ and $\epsilon$ with gluino contributions \cite{eeprime&epsilon}, see also \cite{eeprime}.
It was noted in Ref.\cite{Masiero:2001rg} that the LR mass insertions of the required size may lead to the charge and 
color breaking minima. However, this is not true in general, i.e. when the Higgs, squark, and slepton
mass parameters are unrelated, see Ref. \cite{Casas:1996de}. }. 
For example, only a small  (${\rm Im}(\delta_{LR}^d)_{21}\sim 10^{-5}$)  mass insertion in the down-sector
is required to  generate the observed value of $\varepsilon'$ \cite{Gabbiani:1996hi}. 
Then $\varepsilon$ can be 
entirely due 
to the mass insertions in the up-sector:
 ${\rm Im}(\delta_{LL}^u)_{21}\sim 10^{-3}$ and ${\rm Re}(\delta_{LL}^u)_{21}\sim 10^{-2}$.
Generally, this does not require large SUSY CP-phases and may be accommodated in the framework
of approximate CP symmetry \cite{Dine:2001ne}, which is motivated by the strong EDM bounds (for a review see \cite{Abel:2001vy}).
Alternatively, the CP-phases can be order one but enter only flavor-off-diagonal quantities which
occurs in models with hermitian flavor structures \cite{abel}. 
Clearly, the regions of the parameter space where CP violation can be fully supersymmetric are 
significantly enlarged if both the gluino and the chargino contributions are included.
Of course, it remains a challenge
to build a realistic well-motivated model with all of the required features.

\end{document}